\documentclass[twocolumn, 10pt, pre, aps, showpacs, reprint, amsmath, amssymb, superscriptaddress, nofootinbib]{revtex4-1}

\usepackage[]{graphicx}

\newcommand{\neff}{n_{\mathrm{eff}}}

\newcommand{\ntot}{n_g}

\newcommand{\lpo}{\ell_\mathrm{geo}}

\newcommand{\lopt}{\ell_\mathrm{opt}}
\newcommand{\loptcalc}{\ell_\mathrm{opt}^\mathrm{calc}}
\newcommand{\FSR}{\lambda_\mathrm{FSR}}

\newcommand{\alphacrit}{\alpha_{\mathrm{crit}}}
\newcommand{\alphabrewster}{\alpha_B}

\newcommand{\gth}{g_{th}} 

\newcommand{\reffig}[1]{\mbox{Fig.~\ref{#1}}}

\newcommand{\refeq}[1]{\mbox{Eq.~(\ref{#1})}}
\newcommand{\refsec}[1]{\mbox{Sec.~\ref{#1}}}

\begin{document}

\title{Localized lasing modes of triangular organic microlasers}
\author{C. Lafargue}
\author{M. Lebental}
\email{lebental@lpqm.ens-cachan.fr}
\affiliation{Laboratoire de Photonique Quantique et Mol{\'e}culaire, CNRS UMR 8537, Institut d'Alembert FR 3242, {\'E}cole Normale Sup{\'e}rieure de Cachan, 61 avenue du Pr\'esident Wilson, F-94235 Cachan, France}
\author{A. Grigis}
\affiliation{Laboratoire Analyse, G\'eom\'etrie et Applications, CNRS UMR 7539, Universit\'e Paris 13, Sorbonne-Paris-Cit\'e, 99 avenue J.~B.~Cl\'ement, F-93430 Villetaneuse, France}
\author{C. Ulysse}
\affiliation{Laboratoire de Photonique et Nanostructures, CNRS UPR20, Route de Nozay, F-91460 Marcoussis, France}
\author{I. Gozhyk}
\author{N. Djellali}
\author{J. Zyss}
\author{S. Bittner}
\affiliation{Laboratoire de Photonique Quantique et Mol{\'e}culaire, CNRS UMR 8537, Institut d'Alembert FR 3242, {\'E}cole Normale Sup{\'e}rieure de Cachan, 61 avenue du Pr\'esident Wilson, F-94235 Cachan, France}

\date{\today}

\begin{abstract}
We investigated experimentally the ray-wave correspondence in organic microlasers of various triangular shapes. Triangular billiards are of interest since they are the simplest cases of polygonal billiards and the existence and properties of periodic orbits in triangles are not yet fully understood. The microlasers with symmetric shapes that were investigated exhibited states localized on simple periodic orbits, and their lasing characteristics like spectra and far-field distributions could be well explained by the properties of the periodic orbits. Furthermore, asymmetric triangles that do not feature simple periodic orbits were studied. Their lasing properties were found to be more complicated and could not be explained by periodic orbits.
\end{abstract}


\maketitle

\section{Introduction}
Two-dimensional (2D) billiards long have been studied as model systems with Hamiltonian dynamics. This is in great part due to their seeming simplicity that contrasts the wealth of different dynamical behaviors that they can exhibit, including integrable, chaotic, pseudointegrable, and mixed dynamics. One interesting class of 2D billiards are polygons, of which triangles are the simplest case. While some classes of triangular billiards are well understood, many unsolved problems remain for triangles of asymmetric shape. These open questions concern, for example, the existence, number, and stability with respect to geometric perturbations of their periodic orbits (POs) \cite{Veech1989, Cipra1995, Gutkin1997, Boshernitzan1998, Kenyon2000, Schwartz2006, Hooper2007}. While the existence of at least one PO, the so-called Fagnano's orbit, is assured for acute triangles, and the existence of POs for obtuse triangles with no angle greater than $100^\circ$ has also been proven \cite{Schwartz2006}, it is not known whether any PO exists at all in an arbitrary obtuse triangular billiard. But even for triangles that are known to have one or more POs, their actual construction is often nontrivial, and even the shortest POs can be quite complicated. The search for POs in triangular billiards hence stays a field of active research.

Two-dimensional billiards are also studied in the context of quantum and wave-dynamical chaos \cite{Gutzwiller1990, StoeckmannBuch2000} to understand the manifestation of ray dynamics in the properties of the corresponding wave-dynamical systems. Early experiments concentrated on microwave and acoustic resonators \cite{Sridhar1991, Stoeckmann1990, Richter1999, Bertelsen1999}, and new interest has arisen with the advent of applications like optical microcavities and -lasers \cite{Schwefel2003, Matsko2009}. In particular, the influence of POs on the spectral and emission properties of microlasers is important in view both of a fundamental understanding of these devices and their applications \cite{Tureci2002a, Gmachl2002, Harayama2003, Lebental2007}.

While many microcavities have circular or deformed circular shape, different types of polygonal microresonators also have been investigated. Examples of such structures include semiconductor \cite{Chang2000, Yoon2007}, organic \cite{Lebental2007, Bogomolny2011} and crystal microlasers \cite{Braun2000}, silicon and silica microresonators \cite{Lee2004a, Li2006, Marchena2008}, vertical-cavity surface-emitting lasers (VCSELs) \cite{Chen2008, Chen2009, Chen2010, Chen2011}, and hexagonal zinc oxide nanocavities and -rods \cite{Nobis2005, Yu2007, Dai2009a}. These studies, however, examined only equilateral polygons, while very few experimental investigations of nonequilateral polygonal resonators have been reported \cite{Kurdoglyan2004, Fonte2012}. Furthermore the scattering properties of triangular and other polygonal structures have been studied \cite{Yin1991, Giannini2007, Lucido2010}. Thus the properties of nonequilateral triangular and polygonal microcavities are poorly understood. In this article we study the lasing characteristics of triangular organic microlasers of different shapes with an emphasis on the role of symmetries and their absence.

A laser resonator is usually designed to confine light on a specific periodic ray trajectory and the active medium is positioned to provide optimal overlap with the corresponding resonant states. In the experiments reported here, the point of view was reversed: The geometry of the cavity was given and defined the various possible POs and the distribution of the active medium, but it was not known \textit{a priori} on which PO, if any, the lasing modes would be based. The laser was essentially left free to decide which PO was being favored. While that PO could often be guessed in advance for simple, symmetric triangles, the question of which specific PO is chosen was particularly interesting in cases where even the shortest POs are long and complicated or the POs are even not known at all. Furthermore, the lasing modes need not necessarily be based on a particular PO. The aim was therefore to understand which parameters determine the dominant PO, to what extent the properties of the lasing modes can be explained by that orbit, and which features of triangular microlasers are beyond simple ray-optical explanations.

The article is organized as follows. Section \ref{sec:triangleDynamics} summarizes the key characteristics of classical triangular billiards and their POs, and \refsec{sec:dielres} treats their implications for dielectric resonators. Section \ref{sec:expTechnique} explains the fabrication of our microlasers and the experimental setup for their characterization. The experimental results for different triangular microlasers are presented in \refsec{sec:expResults}, starting with highly symmetric and well-understood cases and going all the way to completely asymmetric cavities. Section \ref{sec:conclusions} concludes with a summary of the results.

\section{Classical dynamics of triangular billiards} \label{sec:triangleDynamics}
Triangles are the simplest type of polygons and hence can serve in many respects as paradigms for more general polygons. The dynamics of a classical polygonal billiard with $M$ vertices depends on the internal angles $\alpha_{j}$ at the vertices, where $j = 1 \dots M$ is the index of the vertex. A polygonal billiard is called rational if all of its angles can be written as a rational multiple of $\pi$, that is, $\alpha_j = m_j \pi / n_j$, where the $m_j$ and $n_j$ are coprime integer numbers. All other polygons are called irrational. The topology of the phase space of a rational polygon is determined by its genus $G$, which is a function of the $m_j$ and $n_j$ \cite{Richens1981}. For $G = 1$ the phase space has the topology of a torus and the billiard is integrable. The three cases of integrable triangle billiards are the equilateral triangle, the right isosceles triangle, and the triangle with angles $\pi/2$, $\pi/3$, and $\pi/6$, which is an equilateral triangle cut in half \cite{Kenyon2000}. For $G > 1$, the phase space resembles a $G$-handled sphere and its dynamics are said to be pseudointegrable \cite{Richens1981}. Irrational polygons have ergodic dynamics \cite{Kerckhoff1986}. It should be noted that the subset of rational polygons is dense in the set of all polygons, that is, each irrational polygon can be approximated by a rational triangle to arbitrary precision.

\begin{figure}[tb]
\begin{center}
\includegraphics[width = 5 cm]{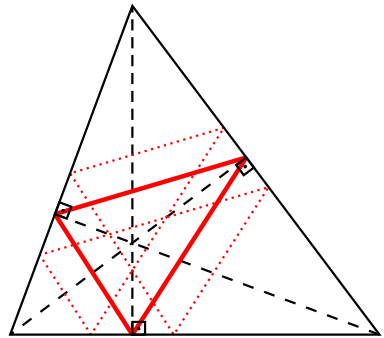}
\end{center}
\caption{(Color online) Fagnano's orbit in an acute triangle. The orbit (thick solid red line) connects the feet of the altitudes (dashed black lines). A trajectory parallel to Fagnano's orbit but starting at a different position closes after two round-trips (dotted red line).}
\label{fig:fagnano}
\end{figure}

\begin{figure*}[tb]
\begin{center}
\includegraphics[width = 14 cm]{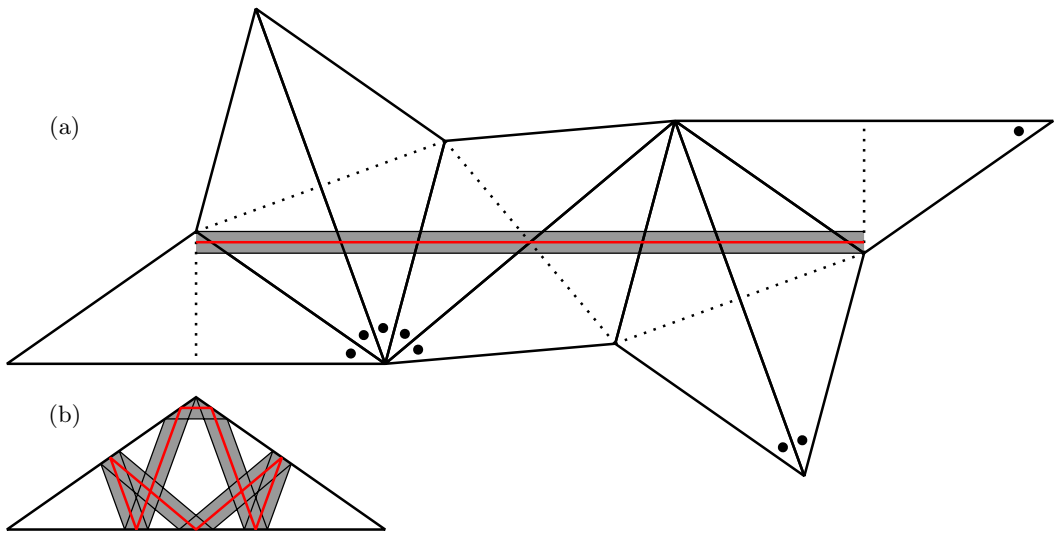}
\end{center}
\caption{(Color online) Double bow-tie orbit in the isosceles triangle with top angle $110^\circ$ (a) unfolded and (b) folded back into the triangle. The red (dark gray) line indicates the isolated PO at the center of the PO channel (gray area) that is restricted by the thin black lines touching the top corner of the triangle. The dotted black lines indicate the height of the triangle and the black dots indicate its orientation.}
\label{fig:dbowtiePO}
\end{figure*}

One type of trajectory that plays an important role in the dynamics of both classical and wave-dynamical billiards are periodic orbits, that is, orbits that retrace themselves after a finite number of reflections. One example is shown in \reffig{fig:fagnano}. While it has been proven that POs exist in any billiard with a smooth ($C^1$) contour \cite{Katok1995}, there is no such theorem for polygonal billiards: It is not known whether any PO exists in an arbitrary polygonal billiard. There are, however, several results for specific cases, one of the oldest concerning acute triangles. It was proven back in the 18th century by Fagnano that the PO connecting the feet of the three altitudes (thick solid red line in \reffig{fig:fagnano}) is the shortest of all possible POs in such triangles \cite{Gutkin1997}. It is hence called Fagnano's orbit. The existence of other POs in irrational acute triangles is, however, not evident.

The POs in rational and several other types of triangles can be constructed with the so-called unfolding technique as demonstrated in \reffig{fig:dbowtiePO}(a) for an isosceles triangle with a top angle of $110^\circ$. We follow a trajectory [red (dark gray) line] that starts perpendicularly to the height of the triangle by reflecting the triangle each time that the boundary is encountered so the trajectory unfolds into a straight line. The trajectory returns to its starting point after a finite number of reflections. The actual PO is obtained by folding the red (dark gray) line back into the triangle as shown in \reffig{fig:dbowtiePO}(b). Furthermore, there are estimates for the number of POs up to a given length in rational triangles, and it has been shown that the POs are dense in their phase space \cite{Boshernitzan1998}. One particular class of rational triangles for which even stronger theorems concerning the POs are known are the so-called Veech triangles \cite{Veech1989, Kenyon2000, Hooper2013}.

Much less is known for irrational triangles. The POs in right and isosceles triangles can also be constructed by unfolding \cite{Cipra1995}. It should be noted, though, that even in cases where POs can be constructed with the unfolding technique, the shortest PO can be long and complicated compared to the simple examples shown in Figs.~\ref{fig:fagnano} and \ref{fig:dbowtiePO}. Furthermore, the existence of POs in triangles with all angles smaller than $100^\circ$ has been proven \cite{Schwartz2006}. However, no general theorems are known concerning the existence of POs in obtuse triangles with one angle larger than $100^\circ$.

All trajectories in polygonal billiards are marginally stable with respect to perturbations of their initial conditions. POs with an odd number of reflections like Fagnano's orbit or the double bow-tie orbit in \reffig{fig:dbowtiePO} are isolated, while POs with an even number of reflections are part of a family of POs with parallel trajectories \cite{Robinett1999, Brack2003}. Hence, repeating an isolated PO an even number of times yields a nonisolated PO. For example, the PO indicated as dotted red line in \reffig{fig:fagnano} belongs to the family of the twice-repeated Fagnano's orbit. The complete family of a PO can be found by unfolding since its other members cover a strip parallel to the PO. This strip is called the PO channel. The PO channel of the double bow-tie orbit in the isosceles $110^\circ$ billiard is indicated as a gray strip in \reffig{fig:dbowtiePO}. The PO channel is bounded by two trajectories (thin black lines) that touch the corners of the triangle. These lines are the optical boundaries of the PO channel. Parallel trajectories beyond these line are not members of that family as can be verified by unfolding them. Depending on the PO and the triangle, the PO channel can cover either a part of the billiard like in \reffig{fig:dbowtiePO} or the complete billiard. For example, the POs of the equilateral and right isosceles triangles cover them completely since these triangles tessellate the plane when unfolding.

\section{Dielectric resonators and periodic orbits} \label{sec:dielres}
The flat organic microlasers studied in this article are treated as 2D passive open dielectric resonators since the lasing modes close to threshold can usually be well described by the modes of the passive cavity. A passive resonator is governed by the scalar Helmholtz equation
\begin{equation} [\Delta + n^2(x, y) k^2] \Psi(x, y) = 0 \end{equation}
where $k$ is the free-space wave number and $n(x, y)$ is the effective refractive index $\neff$ for $(x, y)$ inside the resonator and the refractive index of the surrounding medium (air with $n=1$) on the outside. The wave function $\Psi$ corresponds either to the $z$ component of the electric field, $E_z$, for transverse magnetic (TM) modes or to that of the magnetic field, $B_z$, for transverse electric (TE) modes. The wave functions inside and outside of the resonator are connected by the usual boundary conditions for dielectric interfaces \cite{Jackson1999, Lebental2007, Dubertrand2008}.

Since the typical size of the cavities considered here is in the range of several hundred wavelengths, the resonators are in the so-called semiclassical regime which is the transition regime from classical physics ($\equiv$ ray optics) to quantum mechanics ($\equiv$ wave optics). Semiclassical methods permit to explain various properties of the resonators using concepts and quantities from the dynamics of the corresponding classical billiard systems \cite{Brack2003}, the POs playing an important role in such approximations. Two well-known cases are trace formulas that connect the density of state with the POs \cite{Gutzwiller1970, Gutzwiller1971, Brack2003} and resonant states localized on POs, so-called scars \cite{Heller1984} and superscars \cite{Bogomolny2004, Bogomolny2006}. Trace formulas and scars were mainly investigated in the context of closed resonators with Dirichlet boundary conditions, but the underlying principles can be extended to dielectric resonators. In fact, a trace formula for dielectric resonators has been developed \cite{Bogomolny2008, Bogomolny2011, Bogomolny2012, Bittner2010, Hales2011}, and modes of dielectric resonators localized on classical trajectories are often observed. This includes Gaussian modes that are localized on stable POs \cite{Tureci2002a, Shinohara2011}, scar states localized on unstable POs \cite{Gmachl2002, Harayama2003}, and superscar states localized on families of marginally stable POs \cite{Lebental2007, Song2013a} or classical tori \cite{Bittner2013b}. It has been shown in Ref.~\cite{Bogomolny2003} for resonators with Dirichlet boundary conditions that superscar states are localized inside the PO channel (see \reffig{fig:dbowtiePO}) due to repeated diffraction at the corners that define the optical boundaries. A similar effect has been proposed in Ref.~\cite{Lebental2007} for dielectric resonators even though the diffraction at dielectric corners is not understood \cite{Gennarelli2011}. Therefore we expect to find superscarred lasing modes in triangular dielectric resonators with pseudointegrable classical dynamics.

The influence of a PO on the properties of a resonator depends on several factors, among them its length, its stability or, in the case of nonisolated orbits, the area covered by its family and its refractive losses \cite{Bogomolny1988, Brack2003, Bogomolny2008}. The losses depend on the refractive index and the angles of incidence of a PO since a ray traveling in a dielectric resonator is reflected and refracted at the side walls according to the Fresnel formulas. The emission directions of the refracted rays are determined by Snell's law. The most long-lived modes of passive dielectric cavities hence mainly exhibit the influence of the shortest and best-confined POs \cite{Bittner2010}. On the other hand, modes based on POs that are not confined by total internal reflection can be observed for laser cavities. The threshold condition for a ray traveling along a PO with length $\lpo$ in an active medium with linear gain $g$ is given by
\begin{equation} \exp(g \lpo) \prod \limits_j |r_j|^2 = 1 \end{equation}
where $r_j$ is the Fresnel reflection coefficient for the reflection at the $j$th vertex and the product runs over all vertices of the PO. The threshold gain $\gth$ is hence
\begin{equation} \label{eq:thresGain} \gth = -\frac{2}{\lpo} \sum \limits_j \ln(|r_j|) \, . \end{equation}
We use \refeq{eq:thresGain} as a simple estimate for the threshold of a mode localized on a PO. In practice, however, also other parameters can be of importance like the overlap between the gain region and the mode profiles \cite{Smotrova2011, Bachelard2012} or the coupling between the molecules of the gain medium and the electric field of a mode \cite{Gozhyk2012}. Therefore, we do not expect quantitative agreement of the measured lasing thresholds with \refeq{eq:thresGain}. More sophisticated ap\-proa\-ches are necessary for a quantitative understanding of the lasing thresholds.

It should be noted that even though the POs of a classical billiard can explain many properties of the corresponding resonators, wave effects such as tunneling and diffraction can also have a significant influence. One example is the existence of so-called diffractive orbits that have one vertex at a diffractive corner of the billiard. Diffractive corners are corners with an angle that is not equal to $\pi / m$, where $m$ is an integer. The reflection of a ray at such a corner is not defined in classical mechanics. In contrast, a wave impinging on it is diffracted and hence scattered into various directions. This enables to close ray trajectories that impinge on a diffractive corner and thus diffractive orbits can appear in wave-dynamical billiards (i.e., resonators). For example, they contribute to the trace formula for resonators with diffractive corners or point scatterers \cite{Bruus1996, Bogomolny2000, Hales2011}. Another important point is that the dynamics of wave systems is less sensitive to geometric perturbations than that of classical systems. Therefore the influence of a PO on a wave system can survive geometric perturbations even though the perturbation completely eliminates that PO in the classical dynamics \cite{Bellomo1994, Primack1994}.

\section{Experimental Techniques} \label{sec:expTechnique}
 
\begin{figure}[tb]
\begin{center}
\includegraphics[width = 8.0 cm]{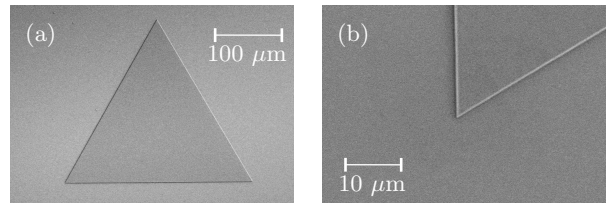}
\end{center}
\caption{SEM images of a microlaser with equilateral triangle shape and $300~\mu$m side length.}
\label{fig:SEMimages}
\end{figure}

The organic microlasers consisted of poly(methyl methacrylate) (PMMA) doped with $5$ wt\% of the laser dye DCM.\footnote{4-(Di\-cyano\-me\-thy\-le\-ne)-2-methyl-6-(4-di\-me\-thy\-la\-mi\-no\-sty\-ryl)-4H-py\-ran (by Exciton)} A solution of PMMA and DCM was spin coated on a silicon wafer with a $2$-$\mu$m-thick layer of silica. The thickness of the PMMA layer was about $700$~nm. The desired cavity shapes were written by $100$-kV electron-beam lithography. This process allowed us to define the cavity boundaries with nanometric precision and achieve vertical side walls and sharp corners and edges \cite{Lozenko2012}. Scanning electron microscope (SEM) images of an equilateral triangle cavity are presented in \reffig{fig:SEMimages}. The triangular microlasers considered here had typical side lengths $a$ in the range of $200$ to $400~\mu$m, i.e., several hundred times larger than the wavelength $\lambda \approx 600$~nm. They are considered two-dimensional (2D) systems with an effective refractive index of $\neff = 1.50$ since they are only about one wavelength thick and support only a single vertical excitation for each polarization \cite{Lebental2007}.

The experimental setup was similar to the one described in Ref.~\cite{Chen2014}. The microlasers were pumped by a pulsed frequency-doubled Nd:YAG laser ($532$~nm, $500$~ps, $10$~Hz, teem photonics PNG-002025-140) that impinged perpendicularly to the cavity plane. The intensity and the polarization of the pump beam were controlled independently using half- and quarter-wavelength plates and polarizers. A circularly polarized pump beam was used, and the pump intensity was normally chosen only slightly above threshold. The pump beam had an approximately Gaussian intensity profile and its diameter was adjusted to cover the complete area of a single microlaser. The lasing emission in the plane of the microlasers was collected in the far field by a lens $18$~cm away from the sample and transferred by a fiber to a spectrometer (Spectra Pro 2500i, Acton Research) and a cooled CCD camera (PIXIS 110B, Princeton Instruments). The spectra were integrated over $10$ pump pulses. The samples could be rotated to record the spectra in all possible directions in the plane of the cavity and thus measure the azimuthal far-field distributions. The polarization of the lasing emission was determined using a linear polarization filter \cite{Gozhyk2012}. All of the microlasers presented in the following emitted transverse electrically (TE) polarized light, i.e., the lasing emission had an electric field parallel to the plane of the cavities. Furthermore, a complementary metal-oxide semiconductor sensor camera (\mbox{UI324xCP-C}, IDS Imaging) with a zoom lens (Zoom 6000, Navitar) was used to take photographs of the lasing cavities. The observation angle of the camera was chosen at a $10^\circ$ tilt angle above the plane so the whole cavities could be surveyed.

The POs that the lasing modes might be localized on can be deduced from the various experimental observables. It should be noted, however, that not all lasing modes are in fact localized on specific classical trajectories. Therefore, a careful analysis of all available data is needed to determine the nature of the observed resonant states. The lasing thresholds are related to the lifetime of the cavity modes and hence to the losses of a possible underlying PO [cf.\ \refeq{eq:thresGain}]. The far-field distributions are often concentrated around a few specific directions. From these directions one can infer the possible trajectories within the resonator via Snell's law. The photographs indicate from which parts of the cavities the light is emitted.

The spectra typically exhibit multimode lasing with several tens of resonances. They are often organized in sequences of equidistant resonances. If a set of lasing modes is localized on a certain PO, their resonance wave numbers are given by
\begin{equation} k_m = \frac{2 \pi m + \theta}{\neff \lpo} \end{equation}
where $m$ is an integer and $\theta$ a constant phase shift. The resonance spacing $k_{m+1} - k_m$ is hence inversely proportional to the optical length $\lopt$ of this PO, which can be conveniently obtained from the Fourier transform (FT) of the spectrum that exhibits peaks at $\lopt$ and its multiples \cite{Lebental2007}. It can also be deduced from the free spectral range (FSR), $\FSR$, via the relation
\begin{equation} \lopt = \frac{\lambda^2}{\FSR} \end{equation}
where $\lambda$ is the wavelength of the lasing emission. The geometric length of the PO, $\lpo$, is related to the optical length by $\lopt = \ntot \lpo$, where $\ntot$ is the group refractive index. The latter differs from the effective refractive index since it also takes into account dispersion. It has a value in the range of $\ntot = 1.60$ to $1.64$ depending on the sample \cite{Lebental2007}. The precise value for each sample can be determined from calibration measurements with ribbon-shaped Fabry-P\'erot cavities since they sustain only a single type of PO, the well-known bouncing ball or Fabry-P\'erot orbits.

\section{Experimental results} \label{sec:expResults}
Seven different triangles with varying degrees of symmetry and different types of classical dynamics were investigated. We start with the simplest and most symmetric triangles and gradually pass to less symmetric and accordingly more complicated ones. The first examples are the equilateral and the right isosceles triangle that have both integrable classical dynamics. The next two triangles are isosceles triangles with top angles $100^\circ$ and $110^\circ$. They are rational triangles with pseudointegrable classical dynamics. Since these four (pseudo-) integrable triangles feature short and simple POs it is expected to find lasing modes localized on some of these orbits. The fifth triangle is an irrational right triangle. It also exhibits a short and simple PO due to its right angle but has ergodic classical dynamics in contrast to the previous examples. Since the rational triangles are a dense subset of all triangles, it is interesting to see if the properties of rational and irrational triangle microlasers significantly differ. The final two triangles are perturbations of the equilateral triangle and the $100^\circ$ isosceles triangle, respectively. They are irrational and hence have ergodic dynamics. They were chosen because no simple POs are known for them except for Fagnano's orbit in the quasiequilateral triangle. In addition, they permit us to study the influence of geometric perturbations on the microlasers' properties and in particular the effect of breaking their mirror symmetry.

\subsection{Equilateral triangle (ET)} \label{ssec:Tequi}
The equilateral triangle (ET) is the triangle with the highest degree of symmetry, and microlasers with equilateral triangular shape have been intensely studied \cite{Chen2008, Chen2009, Chen2010, Chen2011, Chen2003b, Chang2000, Yoon2007, Wysin2005, Wysin2006, Yang2007, Yang2009b, Wang2009}. The equilateral triangle has integrable classical dynamics, and the corresponding cavity problem with Dirichlet or Neumann boundary conditions can be solved analytically \cite{Brack2003}. This problem was already investigated in the context of vibrating membranes by Lam\'e in the 19th century \cite{Lame1852}. There is, however, no analytical solution in the case of the dielectric boundary conditions considered here. All POs of the ET are known, but none of them is confined by total internal reflection. This is due to the relatively low value of $n = 1.5$ that corresponds to a critical angle of $\alphacrit = \arcsin(1 / n) \approx 42^\circ$. The same is true for the other triangular microlasers. The lack of good confinement required comparably large cavities to provide sufficient gain. The side length of the ET microlaser was $a = 300~\mu$m. 

\begin{figure}[tb]
\begin{center}
\includegraphics[width = 8.4 cm]{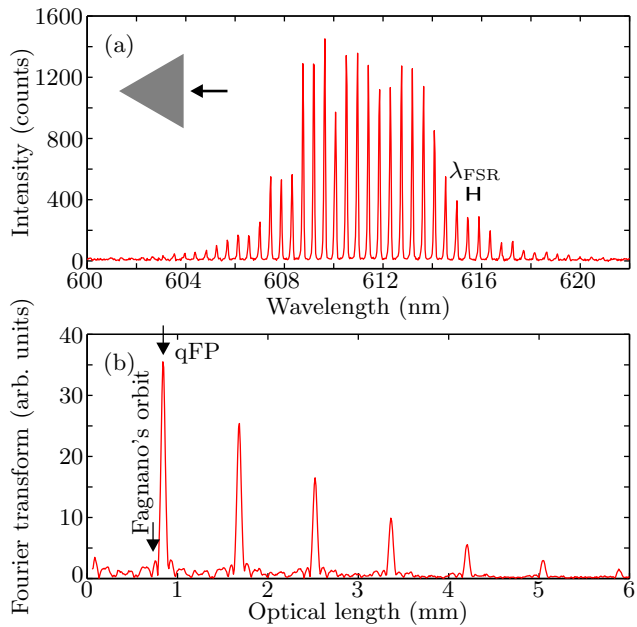}
\end{center}
\caption{(Color online) (a) Spectrum of the equilateral triangle microlaser in the direction $\varphi = 0^\circ$ (as indicated in the inset). (b) Fourier transform of the spectrum. The two arrows indicate the optical lengths of Fagnano's orbit and the quasi-Fabry-P\'erot orbits.}
\label{fig:TequiSpectrum}
\end{figure}

\begin{figure}[tb]
\begin{center}
\includegraphics[width = 7.0 cm]{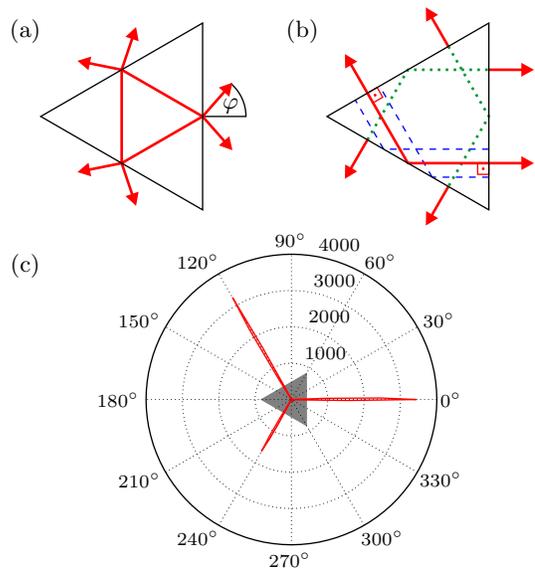}
\end{center}
\caption{(Color online) (a) Fagnano's orbit in the equilateral triangle billiard. The arrows outside the billiard indicate the corresponding emission directions. The azimuthal angle $\varphi$ is the angle with respect to the horizontal axis. (b) Several members of the family of the quasi-Fabry-P\'erot orbit (solid red, dashed blue, and dotted green lines). The arrows indicate the corresponding emission directions. (c) Measured far-field distribution of the equilateral triangle microlaser. The maximal intensity of the spectrum is plotted with respect to the azimuthal angle $\varphi$. The gray triangle in the center indicates the orientation of the cavity.}
\label{fig:TequiFarfield}
\end{figure}

Figure \ref{fig:TequiSpectrum}(a) shows the lasing spectrum $I(\lambda)$ of the ET microlaser for a pump energy just above the threshold in the direction $\varphi = 0^\circ$ [see \reffig{fig:TequiFarfield}(a) for the definition of the azimuthal angle]. The spectrum exhibits a clear structure of equidistant peaks. The Fourier transform of the spectrum, $|\mathrm{FT}(I)|$, is plotted with respect to the optical length $\lopt$ in \reffig{fig:TequiSpectrum}(b). It features several equidistant peaks with decreasing amplitude as expected for a series of equidistant resonances. The first peak at $\lopt = 843~\mu$m corresponds to the FSR, $\FSR = 0.44$~nm, of the lasing spectrum, and the further peaks are harmonics. The two shortest types of POs in the ET billiard are Fagnano's orbit with $\lpo = 3 a / 2 = 450~\mu$m [see \reffig{fig:TequiFarfield}(a)] and the so-called quasi-Fabry-P\'erot (qFP) orbits shown in \reffig{fig:TequiFarfield}(b). A qFP orbit has two reflections with perpendicular incidence and two reflections with an angle of incidence of $60^\circ$ with respect to the surface normal. It has a length of $\lpo = \sqrt{3} a = 519.6~\mu$m. The family of the qFP orbits in the ET is constructed as follows: First, one qFP orbit [e.g., the solid red line in \reffig{fig:TequiFarfield}(b)] is shifted perpendicularly to its trajectory (yielding, for example, the dashed blue lines) and, second, the qFP orbits can be rotated by $\pm 120^\circ$ (yielding, for example, the dotted green lines). The family of qFP orbits covers the whole surface of the ET. The calculated optical lengths corresponding to Fagnano's orbit and the qFP orbit are $\loptcalc = 729~\mu$m and $\loptcalc = 842~\mu$m, respectively. They are indicated by the arrows in \reffig{fig:TequiSpectrum}(b). Obviously, the observed optical length corresponds to the qFP orbit while Fagnano's orbit is too short. All other POs are significantly too long.

\begin{figure}[tb]
\begin{center}
\includegraphics[width = 5.0 cm]{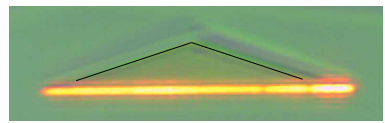}
\end{center}
\caption{(Color online) Photograph of the lasing equilateral triangle microlaser with side length $a = 300~\mu$m in the direction $\varphi = 0^\circ$. The black lines indicate the two other side walls of the cavity.}
\label{fig:TequiPhoto}
\end{figure}

Further evidence is gained from the azimuthal far-field distribution in \reffig{fig:TequiFarfield}(c). The lasing emission is concentrated in the three directions perpendicular to the cavity side walls. This is the behavior expected from modes localized on the qFP orbit as shown in \reffig{fig:TequiFarfield}(b). Note that there is no emission from the reflections with angle of incidence $60^\circ$ since this angle is larger than the critical angle. On the other hand, Fagnano's orbit would correspond to six emission directions with an angle of $\varphi = 48.6^\circ$ with respect to the surface normals as indicated in \reffig{fig:TequiFarfield}(a), but no emission was found in these directions. It should be furthermore noted that the lasing threshold of the ET microlaser is only about $20\%$ higher than that of a Fabry-P\'erot (FP) cavity of corresponding width. Altogether this proves that the observed lasing modes are localized on the qFP orbit. Finally, a photo taken from the direction $\varphi = 0^\circ$ and presented in \reffig{fig:TequiPhoto} shows that the whole side wall of the cavity is lasing. This was expected since the family of the qFP orbit covers the whole triangle, though also all other PO families do so. Photos taken from $\varphi = 120^\circ$ and $240^\circ$ show the same behavior, whereas no lasing light was observed with the camera in all other directions. In summary, the qFP orbit was identified unambiguously from the experimental data as the orbit supporting the lasing modes.

It is at first surprising that the dominant lasing modes are localized on the qFP orbit and not on Fagnano's orbit like in Refs.~\cite{Chang2000, Yoon2007}. First, however, the refractive index of the semiconductor materials used in Refs.~\cite{Chang2000, Yoon2007} was significantly higher so Fagnano's orbit was confined by total internal reflection, which is not the case here. Second, the lasing modes that we observed were TE polarized, i.e., their electric field was parallel to the plane of the resonator. In fact, TE polarized modes are favored by the properties of the lasing dye and the pumping scheme that is used here \cite{Gozhyk2012}. Since the angle of incidence of Fagnano's orbit, $30^\circ$, is close to the Brewster angle $\alphabrewster = \arctan(1 / n) = 33.7^\circ$, a TE mode localized on this PO would suffer from very high losses. From \refeq{eq:thresGain} we calculate $\gth = 359~\mathrm{cm}^{-1}$ as the threshold of Fagnano's orbit and $\gth = 124~\mathrm{cm}^{-1}$ for the qFP orbit. Thus, the dominance of the qFP modes can be well explained by taking into account the peculiarities of the organic microlasers. It should be noted that the modes of the other triangular microlasers considered in the following were all TE polarized as well.

\subsection{Right isosceles triangle (RIT)} \label{ssec:Tdsq}

\begin{figure}[tb]
\begin{center}
\includegraphics[width = 4.0 cm]{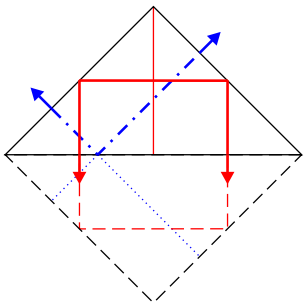}
\end{center}
\caption{(Color online) Right isosceles triangle with the quasidiamond orbit (thick solid red line) and the qFP orbit (dash-dotted blue line). The arrows indicate the emission directions of these two POs. The thin solid red line is the PO along the height of the triangle. The dashed black line indicates the corresponding square billiard and the dashed red line and the dotted blue line indicate the continuation of the quasidiamond and qFP orbit in it, respectively.}
\label{fig:TdsqGeom}
\end{figure}

The second triangle that was investigated is the right isosceles triangle (RIT), which is essentially a square cut in half along a diagonal. The RIT billiard is integrable like the square and equilateral triangle billiards. Both classical and quantum right triangle billiards have been studied and their POs investigated \cite{Cipra1995, Gorin2001, Hooper2007}. A property well known by opticians is the fact that a corner with a right angle sends a ray back parallel to its initial direction regardless of the angle of incidence. From this follows directly the existence of a family of POs that impinge perpendicularly on the hypotenuse as indicated by the thick solid red line in \reffig{fig:TdsqGeom}. We call this orbit quasidiamond orbit in analogy to the diamond PO in the square billiard, indicated by the dashed red line in \reffig{fig:TdsqGeom}. Another important PO is indicated by the dash-dotted blue line. It is reflected perpendicularly at the two short sides of the RIT and with an angle of incidence of $45^\circ$ at the hypotenuse. Since it corresponds to the Fabry-P\'erot orbit of the square billiard (indicated by the dotted blue lines in \reffig{fig:TdsqGeom}), it is also called the qFP orbit. These are the two shortest POs of the RIT, and both PO families cover the whole area of the billiard. The qFP orbit exists in all isosceles triangles, with the angle of incidence on the long side and the area covered by its family depending on the top angle of the triangle.

\begin{figure}[tb]
\begin{center}
\includegraphics[width = 8.4 cm]{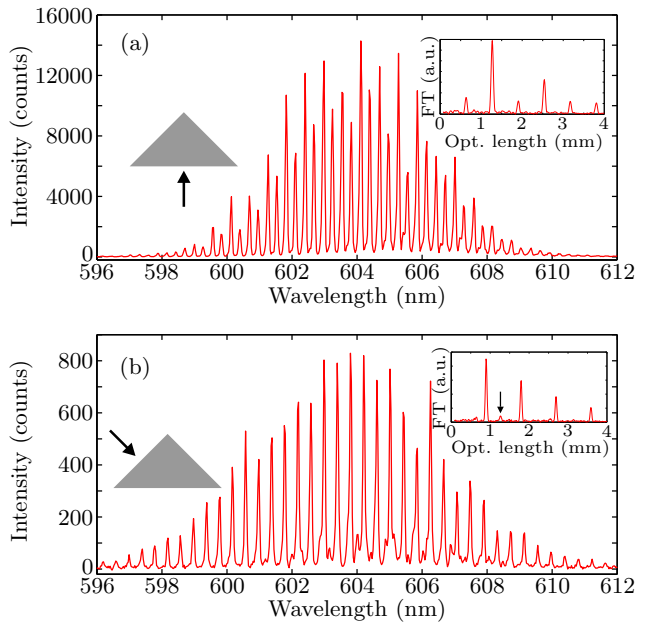}
\end{center}
\caption{(Color online) (a) Spectrum of the right isosceles triangle microlaser measured in the direction $\varphi = 270^\circ$ (see left inset). The right inset shows the FT of the spectrum. (b) Spectrum of the right isosceles triangle microlaser measured in the direction $\varphi = 135^\circ$ (see left inset). The right inset shows the FT of the spectrum, with the arrow indicating a small peak at $\lopt = 1270~\mu$m.}
\label{fig:TdsqSpectraFT}
\end{figure}

Two spectra measured in the directions perpendicular to the hypotenuse and one of the short sides, respectively, are shown in \reffig{fig:TdsqSpectraFT}. Both spectra consist of families of equidistant resonances that have, however, a different FSR each. The largest peak in the FT of the spectrum for $\varphi = 270^\circ$ is at $\lopt = 1270~\mu$m. As expected for this direction, the corresponding PO is the quasidiamond orbit with a geometric length of $\lpo = 2 a = 789.6~\mu$m, where $a = 394.8~\mu$m is the length of the hypotenuse, and a calculated optical length of $\loptcalc = 1263~\mu$m. The FT also features a smaller peak at half this optical length. It stems from the slight modulation of the resonance amplitudes [see \reffig{fig:TdsqSpectraFT}(a)], i.e., the fact that every second resonance has a somewhat smaller amplitude than its neighbors. It is interesting to note that there is an isolated PO along the height of the triangle, indicated as thin solid red line in \reffig{fig:TdsqGeom}, that has half the length of the quasidiamond orbit. The physical origin of the modulation of the resonance amplitudes and whether it is connected to this PO remains, however, unclear. In contrast, the FT of the spectrum at $\varphi = 135^\circ$ shows a peak at $\lopt = 894~\mu$m. This corresponds to the qFP orbit with a geometric length of $\lpo = \sqrt{2} a = 558.3~\mu$m and an optical length of $\loptcalc = 893~\mu$m. An unexpected find is a small peak at $\lopt = 1270~\mu$m [indicated by the black arrow in the inset of \reffig{fig:TdsqSpectraFT}(b)] that corresponds to the quasidiamond orbit. It originates from another family of barely visible resonances in the spectrum.

\begin{figure}[tb]
\begin{center}
\includegraphics[width = 8.4 cm]{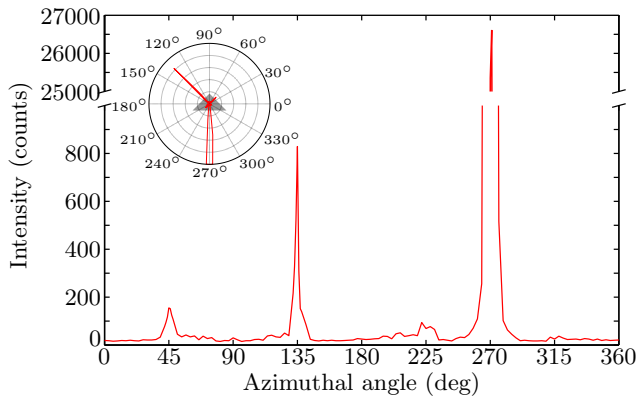}
\end{center}
\caption{(Color online) Measured far-field distribution of the right isosceles triangle microlaser. The maximal intensity of the spectrum is plotted with respect to the azimuthal angle $\varphi$. The inset shows the far-field distribution in polar coordinates where the gray triangle in the center indicates the orientation of the cavity.}
\label{fig:TdsqAzim}
\end{figure}

The far-field distribution shown in \reffig{fig:TdsqAzim} features several emission lobes with differing amplitudes. The strongest emission lobe is in the direction of $\varphi = 270^\circ$ and is due to the quasidiamond orbit. It was cut off in the inset of \reffig{fig:TdsqAzim} since its amplitude of $\approx 26,500$ counts far exceeds that of the other emission lobes. It should be noted that the emission lobe at $135^\circ$ is about $5$ times larger than that at $45^\circ$, whereas the two lobes are expected to have equal amplitudes due to the mirror symmetry of the triangle. A significant asymmetry of the microcavity itself was excluded. Further experiments demonstrated that the ratio between the amplitudes at $\varphi = 45^\circ$ and $135^\circ$ depended sensitively on how precisely the cavity was pumped, that is, for example, on the size and position of the pump beam.

\begin{figure}[tb]
\begin{center}
\includegraphics[width = 7.0 cm]{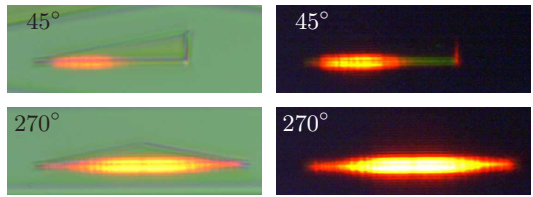}
\end{center}
\caption{(Color online) Photographs of the right isosceles triangle microlaser with a $394.8$-$\mu$m-long hypotenuse taken from $45^\circ$ (top panels) and $270^\circ$ (bottom panels) with background illumination (left panels) and without (right panels).}
\label{fig:TdsqPhotos}
\end{figure}

According to \reffig{fig:TdsqGeom}, modes localized on the qFP orbit should emit in the directions of $45^\circ$ and $135^\circ$, and indeed the spectra measured in these directions exhibit a FSR corresponding to its optical length. However, small contributions also of the quasidiamond orbit were found in the spectra at $45^\circ$ and $135^\circ$ as demonstrated in \reffig{fig:TdsqSpectraFT}(b). In addition, two small lobes at $\varphi = 225^\circ$ and $315^\circ$ were found that are also related to the quasidiamond orbit. The origin of these can be elucidated by the photos shown in \reffig{fig:TdsqPhotos}. The photos taken from $\varphi = 45^\circ$ show a strong emission from the side wall perpendicular to the camera perspective as predicted for the qFP orbit. Classically, it is expected that the whole side wall emits like in \reffig{fig:TequiPhoto} since the qFP orbit family covers the whole triangle. Why this is not observed experimentally remains unclear. It could also be related to the strong sensitivity of the emission to the pumping conditions. In addition, a weak emission from the side wall parallel to the camera perspective was observed. This is best seen in the top right panel of \reffig{fig:TdsqPhotos}. A similar grazing emission was also observed at $\varphi = 135^\circ$, $225^\circ$, and $315^\circ$. It is not expected classically since the quasidiamond orbit is totally reflected at the two short sides. The same kind of grazing emission is also observed for the diamond orbit modes of square organic microlasers and will be discussed elsewhere \cite{LafarguePrep}. The photo taken at $\varphi = 270^\circ$ finally shows emission from the whole hypotenuse as expected for modes localized on the quasidiamond orbit.

The key characteristics of the RIT microlaser can be well explained by simple POs of the corresponding billiard as in the case of the equilateral triangle. But in contrast to the equilateral triangle, the RIT microlaser features two families of modes localized on different POs that coexist. A calculation of the thresholds according to \refeq{eq:thresGain} reveals that they are very close to each other since both POs have the same losses and nearly the same lengths. This prediction agrees qualitatively with the measured thresholds. Another interesting observation is the grazing emission of the modes localized on the quasidiamond orbit that is not expected from the ray dynamics. It demonstrates that some properties of the triangular microlasers need a more careful treatment taking into account wave-dynamical effects.

\subsection{Isosceles triangle with top angle $100^\circ$ (IT100)} \label{ssec:Tiso100}
The most general class of triangles with a symmetry are isosceles triangles. Microlasers with an isosceles triangle shape have been investigated, for example, in Ref.~\cite{Kurdoglyan2004}. The POs of isosceles triangles can be constructed by the unfolding technique \cite{Cipra1995}. The simplest PO that exists in all isosceles triangles is the qFP orbit already known from the right isosceles triangle. Another, more complicated example is the double bow-tie orbit shown in \reffig{fig:dbowtiePO}. It exists for top angles between $90^\circ$ and $111.5^\circ$. The first of two obtuse isosceles triangles that are discussed here is the one with top angle $\alpha = 100^\circ$ (abbreviated IT100 in the following). Its classical dynamics is pseudointegrable.

\begin{figure}[b]
\begin{center}
\includegraphics[width = 8.4 cm]{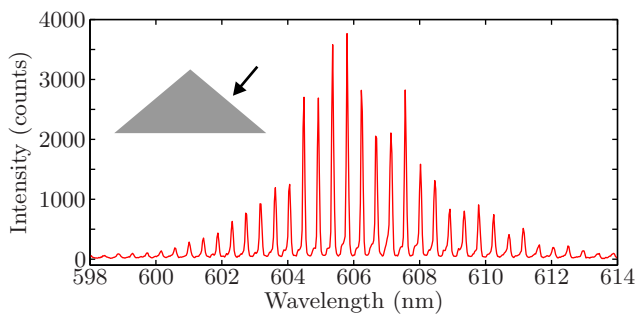}
\end{center}
\caption{(Color online) Spectrum of an isosceles $100^\circ$ triangle microlaser. The inset shows the geometry of the cavity and the angle of observation.}
\label{fig:Tiso100spectrum}
\end{figure}

\begin{figure}[tb]
\begin{center}
\includegraphics[width = 6.0 cm]{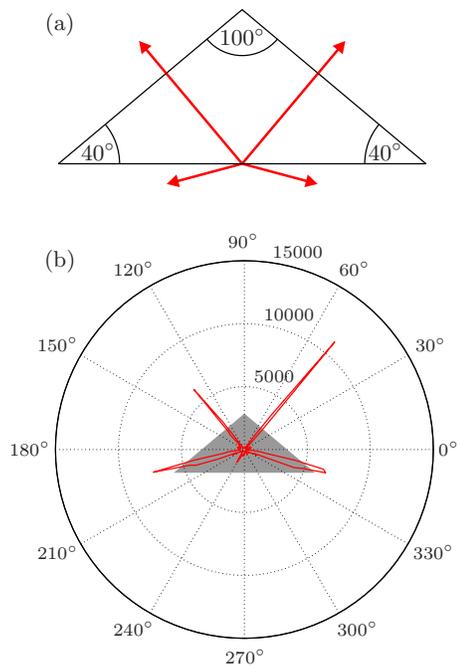}
\end{center}
\caption{(Color online) (a) Geometry of the isosceles $100^\circ$ triangle. A qFP orbit and its emission directions are indicated as red line and red arrows, respectively. (b) Measured far-field distribution of the isosceles $100^\circ$ triangle microlaser. The gray triangle in the center indicates the orientation of the cavity.}
\label{fig:Tiso100Azim}
\end{figure}

The spectrum of the IT100 microlaser measured in the direction $\varphi = 50^\circ$, i.e., perpendicular to one of the short side walls, is shown in \reffig{fig:Tiso100spectrum}. The FSR of the equidistant resonance family corresponds to an optical length of $\lopt = 833~\mu$m. The underlying PO is hence the qFP orbit shown in \reffig{fig:Tiso100Azim}(a) that has a length of $\lpo = 2 a \sin(40^\circ) = 514.2~\mu$m, where the length of the long side is $a = 400~\mu$m, which corresponds to an optical length of $\loptcalc = 833~\mu$m. The far-field distribution presented in \reffig{fig:Tiso100Azim}(b) shows four major emission lobes. Their directions, $\varphi = 50^\circ$, $130^\circ$, $194^\circ$, and $344^\circ$, are precisely the ones expected classically for the qFP orbit. There are also some smaller emission lobes at, e.g., $240^\circ$ and $300^\circ$, hardly visible in \reffig{fig:Tiso100Azim}(b). The spectra in these directions also have a FSR corresponding to the qFP orbit. These directions cannot, however, be explained by the ray dynamics of the qFP orbit.

\begin{figure}[tb]
\begin{center}
\includegraphics[width = 7.5 cm]{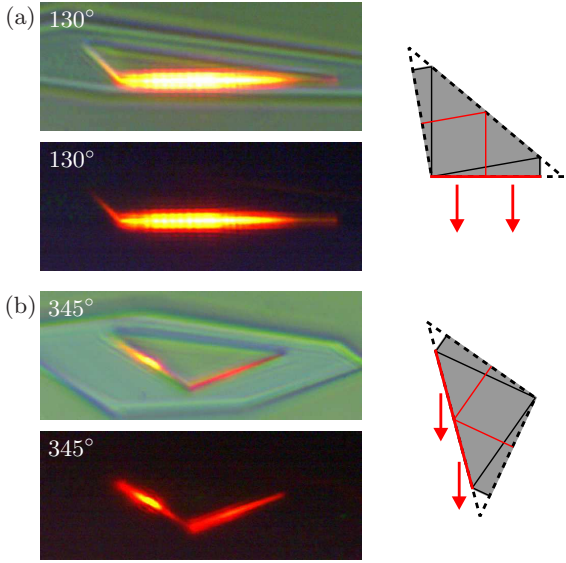}
\end{center}
\caption{(Color online) Photographs of the isosceles $100^\circ$ triangle microlaser (left panels) taken from the directions (a) $130^\circ$ and (b) $345^\circ$ with (top panels) and without background illumination (bottom panels). The right panels show the isosceles $100^\circ$ triangle with the qFP orbit [thin red (dark gray) line]. The solid black lines are the two qFP orbits that define the optical boundaries of the surface covered by the orbit family (gray area), and the parts of the side walls that are accordingly expected to emit in the direction of the arrows are indicated by the thick red (dark gray) lines. The length of the long side is $400~\mu$m.}
\label{fig:Tiso100Photos}
\end{figure}

Figure \ref{fig:Tiso100Photos} presents photographs of the IT100 microlaser taken from two of the major emission directions. The photographs taken from $\varphi = 130^\circ$ in \reffig{fig:Tiso100Photos}(a) show that the whole side wall is emitting with the exception of a small part next to the corner with the long side to the right. This is in fact the part of the boundary that is expected classically to emit, indicated by the thick red (dark gray) line in the sketch in the right panel of \reffig{fig:Tiso100Photos}(a), because the family of the qFP orbit covers only a part of the billiard (indicated as the gray area) in contrast to the right isosceles triangle. The photographs taken from $\varphi = 345^\circ$ in \reffig{fig:Tiso100Photos}(b) demonstrate that the most intense emission into that direction originates from the middle part of the long side (to the left in the photo), again in good agreement with the classical prediction shown in the right panel. However, also the smaller side wall (to the right in the photo) emits light, though with lesser intensity. This emission cannot be explained classically with the properties of the qFP orbit family. In summary, most of the observed lasing characteristics of the IT100 microlaser are in very good agreement with the classical predictions for the qFP orbit, while only some details are beyond a simple ray-dynamical analysis as in the case of the right isosceles triangle microlaser.

\subsection{Isosceles triangle with top angle $110^\circ$ (IT110)} \label{tiso110}

\begin{figure}[tb]
\begin{center}
\includegraphics[width = 8.4 cm]{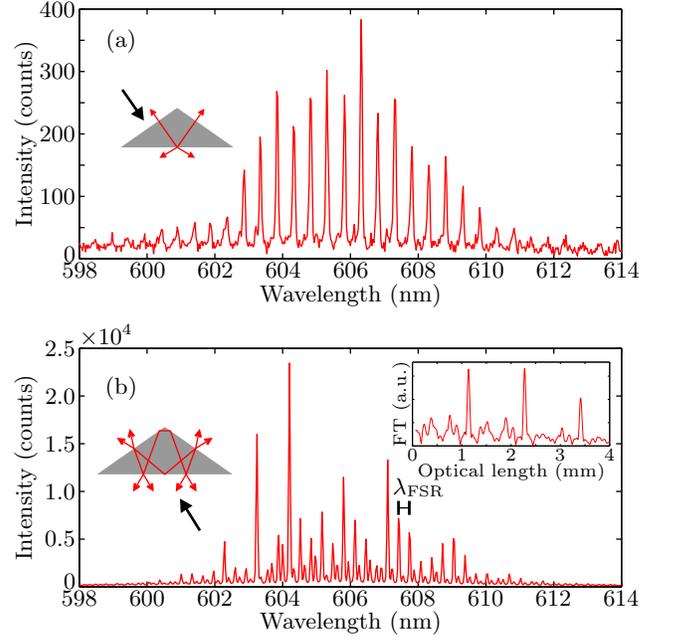}
\end{center}
\caption{(Color online) (a) Spectrum of the isosceles $110^\circ$ triangle microlaser measured at $\varphi = 125^\circ$. The inset indicates the geometry of the triangle, a qFP orbit [red (dark gray) line], and its emission directions [red (dark gray) arrows] as well as the observation direction (black arrow). (b) Spectrum of the isosceles $110^\circ$ triangle microlaser measured at $\varphi = 302^\circ$. The left inset indicates the geometry of the triangle, the double bow-tie orbit [red (dark gray) line], and its emission directions [red (dark gray) arrows] as well as the observation direction (black arrow). The right inset shows the FT of the spectrum.}
\label{fig:Tiso110:spectra}
\end{figure}

The second isosceles triangle that was investigated is the one with a top angle of $110^\circ$ (abbreviated IT110 in the following). It is also pseudointegrable. The spectrum of the IT110 microlaser at $\varphi = 125^\circ$ is shown in \reffig{fig:Tiso110:spectra}(a). It shows a single family of resonances, the FSR of which corresponds to $\lopt = 744~\mu$m. The corresponding PO is again the qFP orbit with geometric length $\lpo = 2 a \sin(35^\circ) = 458.9~\mu$m and optical length $\loptcalc = 743~\mu$m, where the length of the long side is $a = 400~\mu$m. The spectrum measured at $\varphi = 302^\circ$ is presented in \reffig{fig:Tiso110:spectra}(b). Its structure is less clean than that of the one at $125^\circ$, but its FT (see inset) shows clear peaks at $\lopt = 1139~\mu$m and multiples of this length. The corresponding FSR of $\FSR = 0.32$~nm is the FSR of the dominant family of modes [see \reffig{fig:Tiso110:spectra}(b)]. The spectrum also exhibits a second family of modes with smaller amplitude and the same FSR. This optical length as well as the emission direction correspond well to the double bow-tie orbit (shown as inset) with a geometric length of $\lpo = a [1 - \cos(140^\circ)] = 706.4~\mu$m and an optical length of $\loptcalc = 1144~\mu$m. It should be noted that the measured threshold of the modes localized on the qFP orbit is almost $3$ times higher than that of the modes localized on the double bow-tie orbit. This agrees qualitatively with \refeq{eq:thresGain}, which predicts a $2.1$ times higher threshold.

\begin{figure}[tb]
\begin{center}
\includegraphics[width = 8.4 cm]{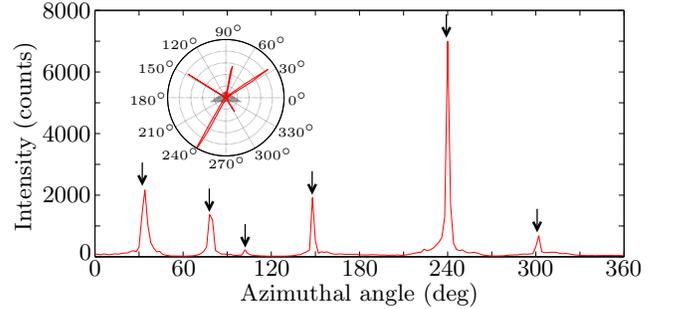}
\end{center}
\caption{(Color online) Experimental far-field distribution of the isosceles $110^\circ$ triangle microlaser. The predicted emission directions of the double bow-tie orbit are indicated by the black arrows. The gray triangle in the inset indicates the orientation of the cavity.}
\label{fig:Tiso110Azim}
\end{figure}

\begin{figure*}[tb]
\begin{center}
\includegraphics[width = 12.0 cm]{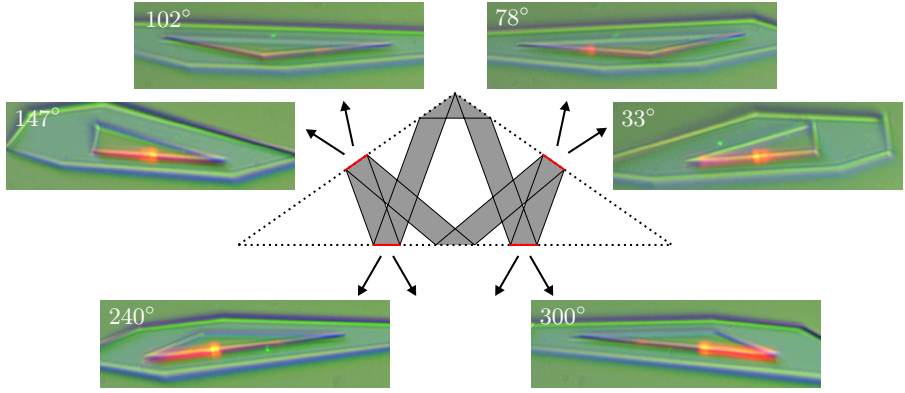}
\end{center}
\caption{(Color online) Photographs of the isosceles $110^\circ$ triangle microlaser taken from the main emission directions of the modes localized on the double bow-tie orbit. The length of the long side is $a = 400~\mu$m. The sketch in the center indicates the area in the isosceles $110^\circ$ triangle that is covered by the family of the double bow-tie orbit (gray area) and the parts of the side walls that are accordingly expected to emit [red (dark gray) lines]. The solid black lines indicate the two POs that form the optical boundaries.}
\label{fig:Tiso110Photos}
\end{figure*}

The far-field distribution of the IT110 microlaser is presented in \reffig{fig:Tiso110Azim}. The six principal emission directions agree very well with those calculated for the double bow-tie orbit that are indicated by the black arrows [see also inset of \reffig{fig:Tiso110:spectra}(b)]. The amplitudes of the emission lobes lack, however, the expected symmetry as in the case of the right isosceles triangle microlaser. No emissions lobes corresponding to the qFP orbit were observed since the microlaser was pumped slightly above the threshold of the double bow-tie orbit modes but well below the threshold of the qFP orbit modes. The photos shown in \reffig{fig:Tiso110Photos} were taken with the same pump intensity as for the measurement of the far-field distribution. The directions of observation correspond to the major emission directions. All photos show that the most intense part of the laser emission originates from those parts of the side walls that are covered by the family of the double bow-tie orbit [indicated by the red (dark gray) lines in the drawing of the IT110]. This confirms that the lasing modes with the lowest threshold are localized on the double bow-tie orbit. However, the photos also show weak emission from other parts of the side walls that are not covered by the family of the double bow-tie orbit. A possible explanation is that even though the field distributions of superscarred resonant states are strongly concentrated inside the PO channel, they also have a nonvanishing field outside of the PO channel due to coupling to nonscarred states \cite{Bogomolny2006, Aberg2008, Dietz2008b}.

In conclusion, the IT110 microlaser is another example like the right isosceles triangle where two families of modes localized on different POs coexist. In contrast to the case of the right isosceles triangle, however, there is a significant difference between the thresholds of the two mode families so one of them is easily selected by keeping the pump intensity sufficiently low. Furthermore, it should be noted that the dominant PO, the double bow-tie orbit, is selected because it has the lowest losses even though it is longer and more complicated than the qFP orbit. The double bow-tie orbit also exists in the IT100 billiard but has a higher threshold than the qFP orbit in that case. Lasing modes based on the double bow-tie orbit were not found experimentally for the IT100 microlasers even at considerably higher pump energies. This demonstrates that the change of a single geometric parameter, the top angle in this case, can significantly modify the lasing characteristics.

\subsection{Pythagorean triangle (PT)} \label{ssec:T345}

\begin{figure}[tb]
\begin{center}
\includegraphics[width = 6.0 cm]{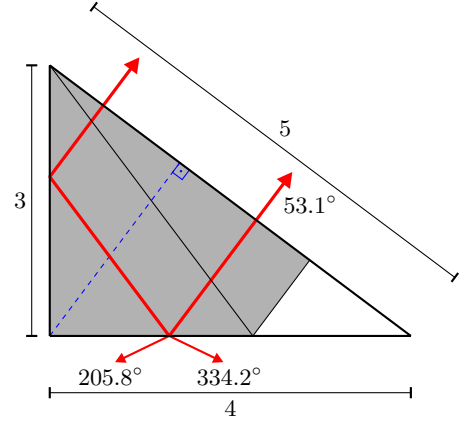}
\end{center}
\caption{(Color online) Pythagorean triangle with side lengths having the ratio $3:4:5$. The thick red (dark gray) line and arrows indicate an example of the qFP orbit family and the corresponding emission directions. The gray area is the surface covered by the qFP orbits that is bounded by the limit orbit indicated as thin black line. The isolated PO along the height of the triangle is indicated as a dashed blue line.}
\label{fig:T345geom}
\end{figure}

The fifth triangle that was investigated is a Py\-tha\-go\-rean triangle (PT) with side length ratio $3:4:5$ as shown in \reffig{fig:T345geom}. Its hypotenuse is $a = 5 \cdot 75~\mu\mathrm{m} = 375~\mu$m long. In contrast to the other triangles studied so far, it is irrational and exhibits no symmetry. It exhibits a qFP orbit that impinges perpendicularly on the hypotenuse as shown in \reffig{fig:T345geom} in analogy to the right isosceles triangle. The POs of right triangles can also be constructed by unfolding \cite{Cipra1995} even if they are irrational like the PT.

\begin{figure}[tb]
\begin{center}
\includegraphics[width = 8.4 cm]{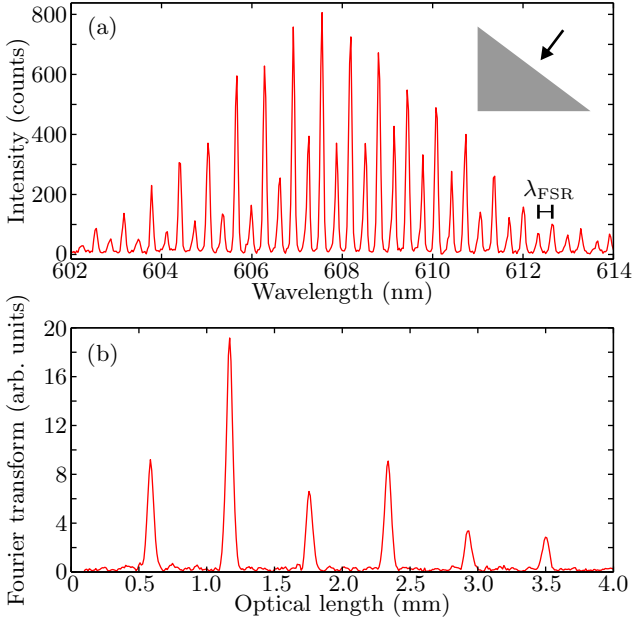}
\end{center}
\caption{(Color online) (a) Spectrum of the Pythagorean triangle microlaser. The inset indicates the direction of observation, $\varphi = 53^\circ$. (b) Fourier transform of the spectrum.}
\label{fig:T345spectrum}
\end{figure}

The lasing spectrum of the corresponding PT microlaser observed in the direction perpendicular to the hypotenuse is presented in \reffig{fig:T345spectrum}(a). It exhibits a single family of equidistant resonances. Its FT, shown in \reffig{fig:T345spectrum}(b), features a dominant peak at $\lopt = 1170~\mu$m that corresponds to the FSR of the spectrum. The length of the qFP orbit is $\lpo = 48 a / 25 = 720~\mu$m, which yields an optical length of $\loptcalc = 1166~\mu$m that is in good agreement with the optical length observed in the FT. In complete analogy to the case of the right isosceles triangle, there is an additional, smaller peak at half this optical length that stems from a modulation of the resonance amplitudes, and there is a PO along the height of the triangle with half the length of qFP orbit (indicated as dashed blue line in \reffig{fig:T345geom}). It should be noted that the relative amplitude of the peaks at $1170~\mu$m and $585~\mu$m depends sensitively on the pump beam position and other details of the excitation scheme.

\begin{figure}[tb]
\begin{center}
\includegraphics[width = 8.4 cm]{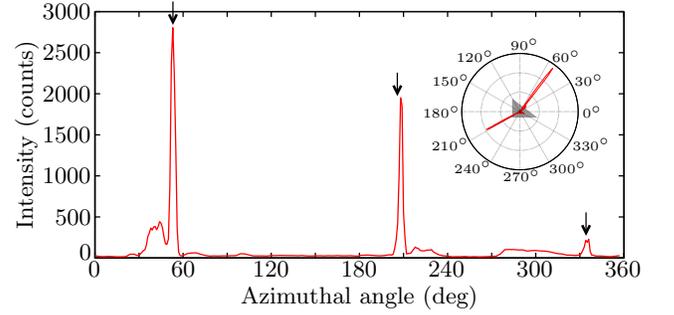}
\end{center}
\caption{(Color online) Measured far-field distribution of the Pytha\-go\-rean triangle microlaser. The emission directions of the qFP orbit are indicated by the black arrows. The gray triangle in the inset indicates the orientation of the cavity.}
\label{fig:T345azim}
\end{figure}

The far-field distribution presented in \reffig{fig:T345azim} has three major emission directions, $\varphi = 53^\circ$, $209^\circ$, and $335^\circ$. The lasing spectra in these directions exhibit the same modes as the one shown in \reffig{fig:T345spectrum}(a). The emission directions predicted for the qFP orbit according to Snell's law for $n = 1.5$ are $\varphi = 53^\circ$, $206^\circ$, and $334^\circ$, respectively (see \reffig{fig:T345geom}). They are indicated in \reffig{fig:T345azim} by the black arrows and agree quite well with the observed ones. It should be noted that no emission is expected from the left (smallest) side of the PT since the angle of incidence of the qFP orbit on it is larger than the critical angle. The amplitude of the emission lobe at $335^\circ$ is considerably smaller than that of the lobe at $209^\circ$ while an equal amplitude is expected classically. In fact, the relative amplitude of the lobes strongly depended on the position and size of the pump beam as in the case of the right isosceles triangle microlaser.

Furthermore, the directions of these two lobes are not symmetric as well. According to the geometry of the qFP orbit, their angles with respect to the surface normal should be equal, but in reality they are $270^\circ - 209^\circ = 61^\circ$ and $335^\circ - 270^\circ = 65^\circ$, respectively, and thus differ by $4^\circ$. This is a significant deviation that is within the resolution of the setup. The third emission lobe on the contrary has precisely the expected direction perpendicular to the hypotenuse. In contrast to the previously considered triangles, however, the PT cavity itself exhibits no symmetry, and symmetric emission directions are only expected due to the properties of the underlying qFP orbit.

A completely unexpected experimental result are the three smaller and broader emission lobes around $\varphi = 40^\circ$, $220^\circ$, and $330^\circ$. The spectra in these directions feature the same structure and FSR as those in the three major emission lobes; however, these three emission directions cannot be related to the dynamics of the qFP orbit.

\begin{figure}[tb]
\begin{center}
\includegraphics[width = 7.5 cm]{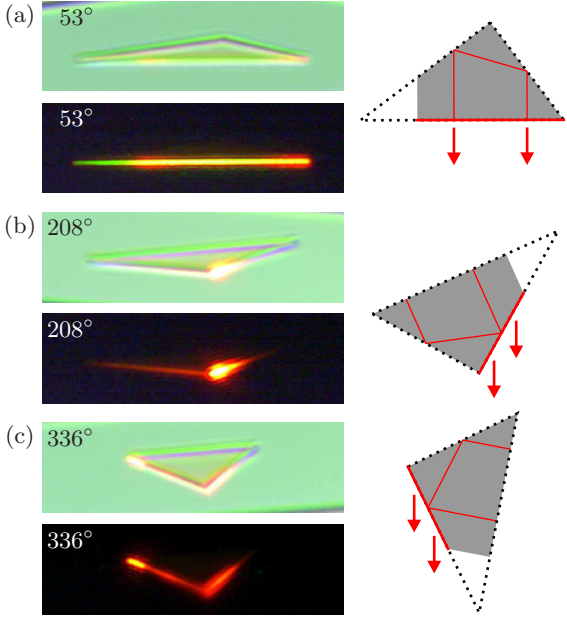}
\end{center}
\caption{(Color online) Photographs of the Pythagorean triangle microlaser (left panels) taken from the directions (a) $\varphi = 53^\circ$, (b) $208^\circ$, and (c) $336^\circ$ with (top panels) and without background illumination (bottom panels). The right panels show the Pythagorean triangle with a qFP orbit [thin red (dark gray) line], the area covered by its family (gray area), and the parts of the side walls that are hence expected to emit [thick red (dark gray) lines] in the direction of the arrows. The length of the hypotenuse is $375~\mu$m.}
\label{fig:T345photos}
\end{figure}

Photographs of the PT microlaser taken from the three major emission directions are shown in the left panels of \reffig{fig:T345photos}. The sketches in the right panels indicate the parts of the side walls that are expected to emit in these directions according to the geometry of the qFP orbits. Indeed, the brightest areas of emission in the photos correspond well to these classical predictions. In contrast, also some weak emissions are observed from the side wall on the left side at $208^\circ$ and from the side wall at the right side at $336^\circ$, which are not expected classically.

In summary, the spectra, the far-field distribution, and the photos clearly evidence that the observed lasing modes are localized on the qFP orbit family. In detail, however, there are deviations from the ray-dynamical predictions. First, the directions of the two emission lobes at $209^\circ$ and $335^\circ$ do not exhibit the expected symmetry; second, there are three additional emission lobes the directions of which have no apparent connection to the qFP orbits; and, third, parts of the lasing emission stem from sections of the cavity boundary that are not expected to emit classically. So even though the underlying PO could be identified, the properties of the PT triangle microlaser cannot be explained with the same precision and completeness as in the previous cases. This might be related to the fact that the PT billiard is neither rational nor symmetric, and in fact its POs can be constructed easily only due to its right angle.

\subsection{Quasiequilateral triangle (QET)} \label{TQequi}

\begin{figure}[tb]
\begin{center}
\includegraphics[width = 5.0 cm]{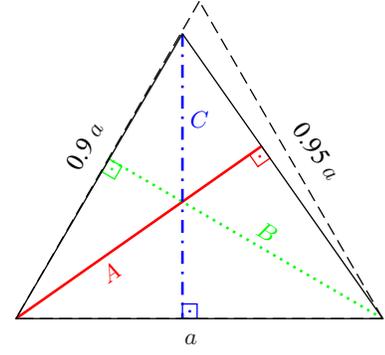}
\end{center}
\caption{(Color online) Geometry of the quasiequilateral triangle. The side lengths are indicated in units of the largest side length $a$. The solid red, dotted green, and dash-dotted blue lines along the heights of the triangle indicate the diffractive POs $A$, $B$, and $C$, respectively. The dashed lines indicate an equilateral triangle with side length $a$ for comparison.}
\label{fig:TQequiGeom}
\end{figure}

While all the triangles considered so far had symmetries or other properties that enabled an easy construction of their POs, the case considered in the following has none of these. It is a deformation of the equilateral triangle and hence called quasiequilateral triangle (QET) in the following. Its side lengths are $a = 316.7~\mu$m, $0.95 a = 308.8~\mu$m, and $0.9 a = 285.0~\mu$m, respectively, as indicated in \reffig{fig:TQequiGeom}. It is an irrational triangle and we hence cannot easily construct any PO besides Fagnano's orbit. In particular, the qFP orbits no longer exist due to the lack of symmetry.

\begin{figure}[tb]
\begin{center}
\includegraphics[width = 8.4 cm]{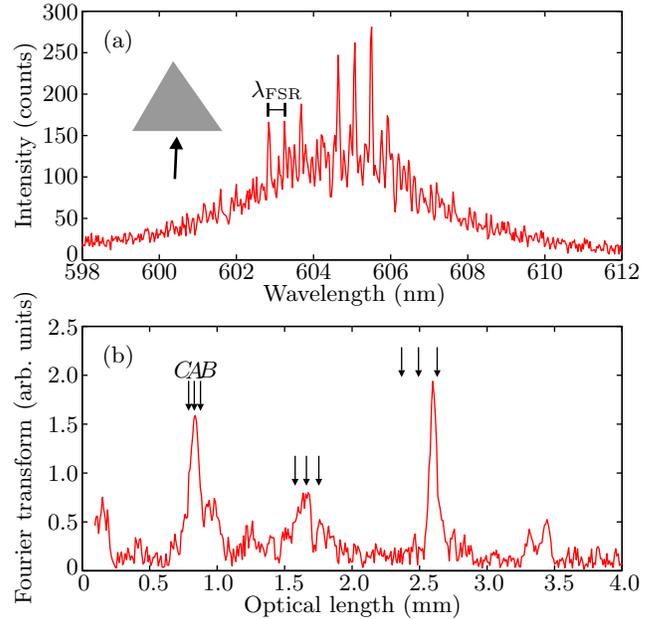}
\end{center}
\caption{(Color online) (a) Spectrum of the quasiequilateral triangle microlaser. The observation angle of $\varphi = 267^\circ$ is indicated in the inset. The indicated FSR corresponds to the dominant peak in the FT of the spectrum. (b) FT of the spectrum. The triplets of arrows indicate the optical lengths of the three diffractive POs $A$, $B$, and $C$ and their multiples.}
\label{fig:TQequiSpectrumFT}
\end{figure}

A typical spectrum, observed at $\varphi = 267^\circ$, is shown in \reffig{fig:TQequiSpectrumFT}(a). The structure of the spectrum is not as clear as for the previously shown ones; nonetheless, a small sequence of equidistant resonances can be identified. The FT of the spectrum shows a peak at $\lopt = 830~\mu$m and its approximate multiples. The corresponding FSR of $\FSR = 0.44$~nm matches that of the resonance sequence in the spectrum. This optical length is close to that of the qFP orbit in an equilateral triangle with side length $a$, but the qFP orbit no longer exists in the QET. There are, however, three diffractive POs along the heights of the triangle that have nearly the same lengths. They are depicted in \reffig{fig:TQequiGeom} and called $A$, $B$, and $C$ in the following. They are called diffractive orbits because one of their vertices is at a diffractive corner of the billiard (cf.\ \refsec{sec:dielres}). It should be noted that for an equilateral triangle, the orbits along the heights belong to the family of the qFP orbit. Thus, these three diffractive POs can be considered as the remnants of the qFP orbit family. Their lengths are $\lpo^{(A)} = 518.2~\mu$m, $\lpo^{(B)} = 547.0~\mu$m, and $\lpo^{(C)} = 492.2~\mu$m. The corresponding optical lengths are indicated by arrows in \reffig{fig:TQequiSpectrumFT}(b) and are indeed close to the peaks observed in the FT of the spectrum.

\begin{figure}[tb]
\begin{center}
\includegraphics[width = 7.0 cm]{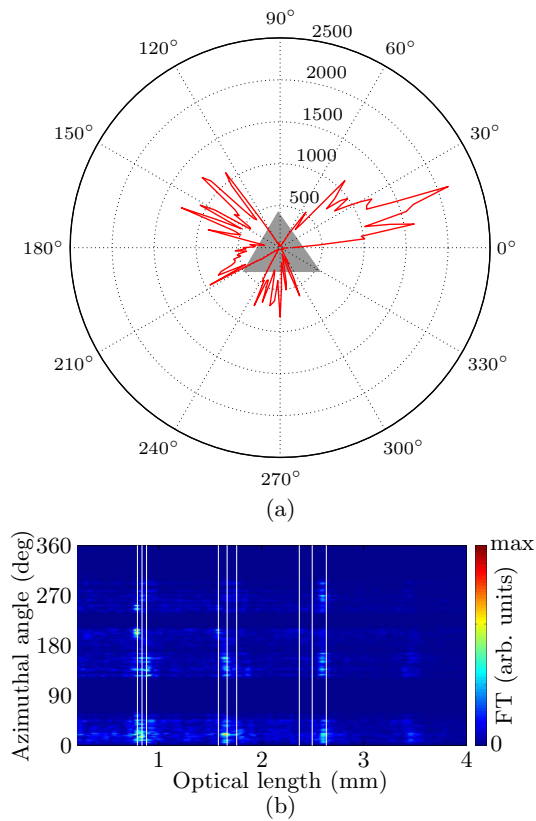}
\end{center}
\caption{(Color online) (a) Measured far-field distribution of the quasiequilateral triangle microlaser. The gray triangle in the center indicates the orientation of the cavity. (b) FT of the spectra with respect to the optical length and azimuthal angle $\varphi$. The vertical white lines indicate the optical lengths of the three diffractive POs and their multiples.}
\label{fig:TQequiAzim}
\end{figure}

The far-field distribution of the QET microlaser is plotted in \reffig{fig:TQequiAzim}(a). It features three broad bundles of emission lobes the centers of which are roughly perpendicular to the cavity side walls. The FT of the spectra measured at different azimuthal angles is shown in \reffig{fig:TQequiAzim}(b). The dominant optical lengths vary somewhat with $\varphi$, but they always stay close to the lengths of the three diffractive POs, the optical lengths of which are indicated by the vertical white lines. This corresponds to the fact that the spectra always feature a similar FSR even though their structure and quality varies significantly, leading to a relatively high noise level in the FT. No evidence of Fagnano's orbit was found like in the case of the equilateral triangle since its angles of incidence are close to the Brewster angle. So while the FSRs approximately match those corresponding to the diffractive orbits, other observations do not indicate that the modes are localized on them. For example, the directions of maximal emission are not exactly perpendicular to the side walls as one would naively expect.

\begin{figure}[tb]
\begin{center}
\includegraphics[width = 5.0 cm]{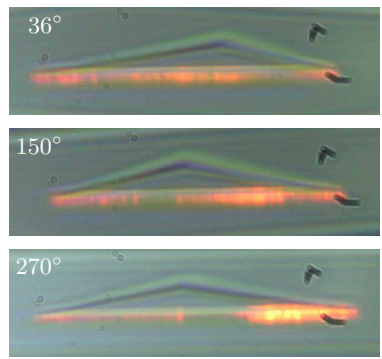}
\end{center}
\caption{(Color online) Photographs of the quasiequilateral triangle microlaser taken from the directions perpendicular to the side walls, $\varphi = 36^\circ$ (top), $150^\circ$ (middle), and $270^\circ$ (bottom). The largest side is $a = 316.7~\mu$m long.}
\label{fig:TQequiPhotos}
\end{figure}

The photos shown in \reffig{fig:TQequiPhotos} enable a better understanding of the nature of the resonant modes. They show the lasing QET from the directions approximately perpendicular to the side walls. All of them show that the points of origin of the lasing emission are more or less broadly distributed over the side walls. The expectation for modes localized on the diffractive POs, in contrast, would be that the origin of emission is strongly concentrated around the feet of the heights that are roughly in the middle of the side walls. Another expectation for this kind of mode would be strong emission coming from the corners of the triangle. This is, however, not observed in any direction. In fact, the images shown in \reffig{fig:TQequiPhotos} are typical also for other directions in which the QET microlaser emits.

In conclusion, the lasing characteristics of the QET microlaser cannot be explained by any PO of the QET billiard. The observed FSRs correspond to an optical length similar to that of the qFP orbits of an equilateral triangle. A possible explanation is that the observed modes are the perturbed modes of the equilateral triangle microlaser that were localized on the qFP orbit. Due to the small perturbation, their FSR stays approximately the same, but the far-field distribution broadens around the emission directions of the qFP orbit that are perpendicular to the side walls. The same effect was observed in Ref.~\cite{Bellomo1994} for triangular resonators with Dirichlet boundary conditions. Some modes were shown to be localized on so-called ghost POs, i.e., the POs of a geometrically different, but similar, triangle. An analogous case is the persistence of the influence of the bouncing ball orbits in a quantum stadium billiard when the originally parallel side walls of the stadium are slightly tilted \cite{Primack1994}. The reason for these effects is that while the classical dynamics can exhibit singular behavior with respect to perturbations of the billiard geometry, e.g., POs suddenly vanishing completely, wave-dynamical systems react in a continuous manner to geometric perturbations, essentially smoothing out the singularities of classical mechanics. The data presented here lead us to believe that the modes of the quasiequilateral triangle microlasers are localized on the ghost qFP orbit, but numerical investigations of the wave functions will be necessary to confirm this notion.

\subsection{Quasi-isosceles triangle (QIT)} \label{sec:TQiso100}

\begin{figure}[tb]
\begin{center}
\includegraphics[width = 6.0 cm]{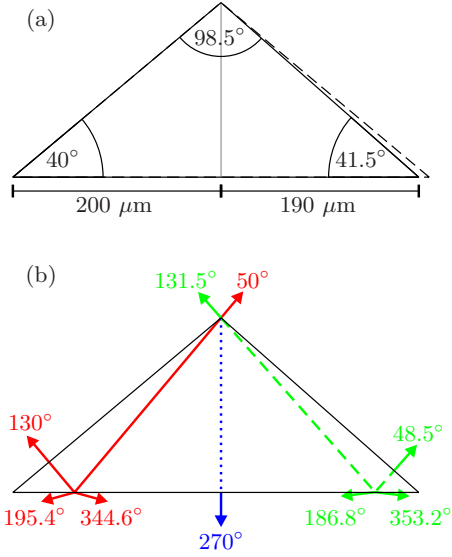}
\end{center}
\caption{(Color online) (a) Geometry of the quasi-isosceles triangle. It is a deformation of the isosceles $100^\circ$ triangle that is indicated by the dashed lines. The gray line indicates the height. (b) Three diffractive POs of the quasi-isosceles triangle. The solid red and dashed green lines indicate the two diffractive qFP orbits, respectively, and the dotted blue line the height orbit. The arrows indicate the corresponding emission directions.}
\label{fig:TQiso100}
\end{figure}

The last triangle is again a deformation of one of the previous triangles. The triangle was constructed by moving the right vertex of the isosceles $100^\circ$ triangle by $10~\mu$m to the left while keeping the other two vertices fixed as demonstrated in \reffig{fig:TQiso100}(a). Hence it is called quasi-isosceles triangle (QIT). It is irrational and therefore no simple POs are known since the qFP and the double bow-tie orbit of the isosceles $100^\circ$ triangle are destroyed by the geometric perturbation. There are, however, three diffractive POs [see \reffig{fig:TQiso100}(b)] similar to the case of the quasiequilateral triangle. One is along the height of the triangle (called the height orbit in the following), and the other two have one perpendicular reflection at a short side wall and one at the top vertex. They can be considered as the remnants of the qFP orbits of the isosceles $100^\circ$ triangle and are called diffractive quasi-Fabry-P\'erot orbits in the following.

\begin{figure}[tb]
\begin{center}
\includegraphics[width = 8.4 cm]{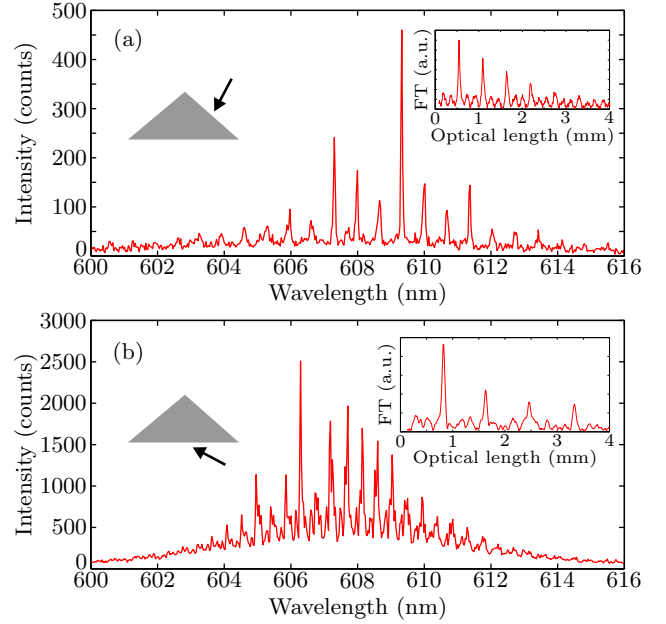}
\end{center}
\caption{(Color online) (a) Spectrum of the quasi-isosceles triangle microlaser observed at $\varphi = 62^\circ$. (b) Spectrum of the quasi-isosceles triangle microlaser observed at $\varphi = 332^\circ$. The insets in (a) and (b) indicate the observation direction and the FT of the spectrum, respectively.}
\label{fig:TQiso100Spectra}
\end{figure}

\begin{figure}[tb]
\begin{center}
\includegraphics[width = 8.4 cm]{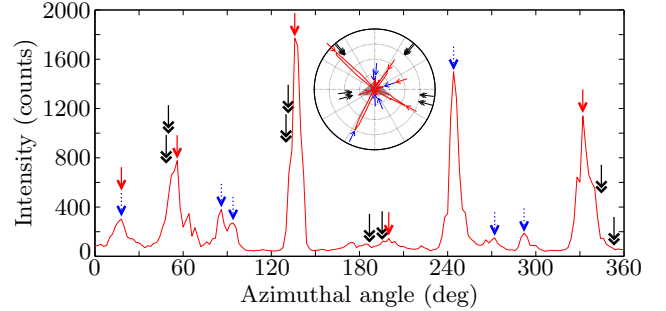}
\end{center}
\caption{(Color online) Measured far-field distribution of the quasi-isosceles triangle microlaser. The gray triangle in the inset indicates the orientation of the cavity. The solid red and dotted blue arrows indicate the directions in which the FT of the spectrum exhibited the optical length of the diffractive qFP and the height orbit, respectively. The black double arrows indicate the emission directions predicted for the diffractive qFP orbits.}
\label{fig:TQiso100Azim}
\end{figure}

Two spectra of the QIT microlaser observed in different directions are plotted in \reffig{fig:TQiso100Spectra}. Both spectra show a relatively clear structure of equidistant modes. However, their FSRs and the corresponding optical lengths differ as can be seen in the FTs shown as insets. The optical length for the spectrum at $\varphi = 62^\circ$ is $\lopt = 546~\mu$m. This can only correspond to the height orbit with $\lpo = 335.6~\mu$m and $\loptcalc = 547~\mu$m. The optical length for the spectrum at $\varphi = 332^\circ$ is $819~\mu$m and corresponds approximately to the lengths of the two diffractive qFP orbits. The geometric and optical length of the left diffractive qFP orbit [solid red line in \reffig{fig:TQiso100}(b)] are $\lpo = 514.2~\mu$m and $\loptcalc = 838~\mu$m, respectively, and are identical to those of the qFP orbit in the isosceles $100^\circ$ triangle, while the geometric and optical length of the right diffractive qFP orbit [dashed green line in \reffig{fig:TQiso100}(b)] are $\lpo = 503.1~\mu$m and $\loptcalc = 820~\mu$m, respectively. So the QIT microlaser exhibits (at least) two different families of modes that have similar thresholds but different FSRs.

Next, we investigated what family of modes emitted in which directions. The far-field distribution in \reffig{fig:TQiso100Azim} shows a large number of emission lobes with varying amplitudes, though the emission is not as broadly distributed as in the case of the quasiequilateral triangle. The directions in which the spectra exhibit the optical length of the diffractive qFP orbits and the height orbit are indicated by the solid red and dotted blue arrows, respectively. Note that for some of the smaller emission lobes the corresponding optical length could not be clearly determined due to indistinct spectra. In fact, some of the most prominent emission lobes point approximately in the directions expected for a diffractive qFP orbit which are indicated by the black double arrows [see also \reffig{fig:TQiso100}(b)]. There remain, however, significant deviations between the emission directions of the presumed diffractive qFP modes and the directions predicted by ray optics. These deviations cannot be consistently explained by a different refractive index either. Regarding the presumed height orbit modes, there are several emission lobes around, though not precisely in, the direction of $270^\circ$ predicted classically. Furthermore, the two emission lobes close to $90^\circ$ seem reasonable for such modes, too. The strong emission in the direction of $\varphi = 244^\circ$ from the presumed height orbit modes and in the direction of $16^\circ$ from both families of modes, however, defy any simple ray-dynamical explanation. Thus, the dynamics of the diffractive POs can explain the observed emission directions at best on a qualitative level.

\begin{figure}[tb]
\begin{center}
\includegraphics[width = 7.0 cm]{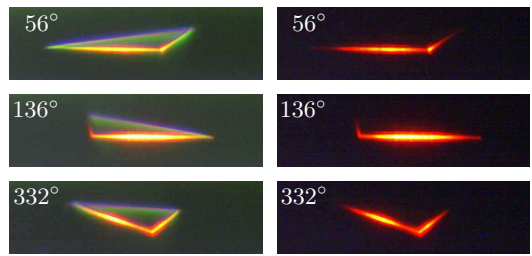}
\end{center}
\caption{(Color online) Photographs of the quasi-isosceles triangle microlaser taken in the main emission directions featuring the optical length of the diffractive qFP orbits, $\varphi = 56^\circ$ (top), $136^\circ$ (middle), and $332^\circ$ (bottom), with background illumination (left panels) and without (right panels). The largest side is $390~\mu$m long.}
\label{fig:TQiso100PhotosFP}
\end{figure}

\begin{figure}[tb]
\begin{center}
\includegraphics[width = 7.0 cm]{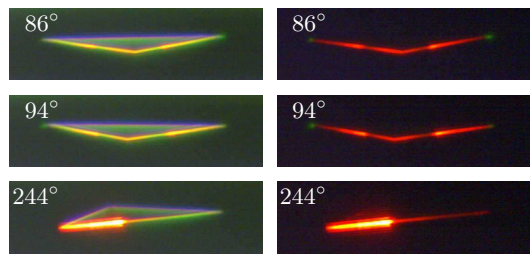}
\end{center}
\caption{(Color online) Photographs of the quasi-isosceles triangle microlaser taken in the main emission directions featuring the optical length of the height orbit, $\varphi = 86^\circ$ (top), $94^\circ$ (middle), and $244^\circ$ (bottom), with background illumination (left panels) and without (right panels).}
\label{fig:TQiso100PhotosHPO}
\end{figure}

The photos taken in the main emission directions of the presumed diffractive qFP modes are presented in \reffig{fig:TQiso100PhotosFP}. They show that the emission originates from almost the whole side walls and not just small portions of it as would be expected for the isolated diffractive qFP orbits. The photos rather resemble those of the equilateral and isosceles triangle microlasers. In addition, a part of the emission originates from those side walls that are not directly facing the camera. Emission from these side walls is not at all expected from any Fabry-P\'erot-like modes. The photos in \reffig{fig:TQiso100PhotosHPO} were analogously taken in the main emission directions of the presumed height orbit modes. The lasing emission is not as broadly distributed along the side walls as in the cases presented in \reffig{fig:TQiso100PhotosFP} but shows nonetheless no single points of concentration. In particular, the photos at $\varphi = 86^\circ$ and $94^\circ$ demonstrate that the emission in these directions is not predominantly originating from the vertex of the triangle as expected for a mode localized on the diffractive height orbit.

In summary, the photos evidence that both families of lasing modes cover more or less the whole microlaser and are not strongly concentrated along the isolated diffractive orbits shown in \reffig{fig:TQiso100}. We are hence led to believe that one family of modes is localized on the ghost qFP orbits, i.e., they are the perturbed qFP modes of the isosceles $100^\circ$ triangle, analogously to the case of the quasiequilateral triangle. The situation for the other family of modes with an FSR corresponding to the height orbit is less clear since a (nondiffractive) height PO does not exist in the isosceles $100^\circ$ triangle, and no family of modes with a similar FSR was observed for the isosceles $100^\circ$ triangle microlaser. In any case, further investigations are necessary for a full understanding of the QIT microlaser's behavior. It is interesting to note that even though a PO is known to exist in the QIT since its angles are all less than $100^\circ$, the dominant lasing modes are clearly not localized on any classical PO. It can be presumed that even the shortest PO in the QIT is too long or has too high losses to support lasing modes with a reasonably low threshold and that therefore other types of lasing modes are predominant.

\section{Conclusions} \label{sec:conclusions}
While there are still open questions regarding the existence and properties of POs in classical triangular billiards and this remains a domain of active research, it is also very interesting to explore the influence of the POs on the properties of the corresponding wave-dynamical billiards. Of particular interest are billiards for which the POs cannot be easily constructed or are not even known because they are irrational and lack symmetries. One of the objectives of the experiments presented here was therefore to see whether the resonant states in these cases are localized on POs, and, if yes, on which ones or if, on the contrary, they have no relation to specific POs. We investigated this ray-wave correspondence in experiments with organic microlasers of triangular shape. The shapes that were chosen correspond to different types of classical dynamics featuring diverse types of POs and were hence expected to exhibit very different lasing characteristics. Even though each triangle microlaser had properties distinct from the others and had to be treated separately, it is possible to draw some general conclusions from the experiments.

Several examples of triangles with well-understood classical dynamics and relatively simple POs were studied, namely the equilateral triangle and several rational isosceles triangles. They all exhibit lasing modes that are clearly localized on POs. The most important observables were the FSR of the spectra and the directions of the emission lobes in the far field that could be measured with high precision and showed excellent agreement with ray-optical predictions. In contrast, the amplitudes of the emission lobes proved to be very sensitive to small changes in the experimental setup. Photographs of the lasing cavities allowed further insight into the characteristics of the lasing modes and showed good qualitative agreement with the classical predictions. It seems to be a general rule for polygonal billiards having simple, short POs with not too high losses that the modes of the corresponding microlasers are localized on these POs \cite{Lebental2007, Bittner2010, Bogomolny2011, Bittner2013b}. Nonetheless, even for these simple and symmetric triangles some details of the observed spectra and far-field distributions were beyond simple ray-dynamical explanations. This underlines the need to further refine the different models of modal localization in dielectric resonators to better account for wave-dynamical effects like diffraction.

The agreement with the ray-optical calculations was less good for the case of an irrational right triangle, namely the Pythagorean triangle. Due to its right angle, it still features a simple PO, and the FSR of the spectra corresponded to its optical length. The directions of the emission lobes, however, showed deviations from the classical expectations, and several emission lobes in addition to the predicted ones were found. Also the photographs of the PT microlaser only partly agreed with the classical expectations.

Finally, two examples of irrational triangles lacking any symmetry and hence any simple POs (with the exception of Fagnano's orbit for the quasiequilateral triangle) were investigated. Both cases were deformations of previously investigated triangles. Their modes seemed to be localized on ghost POs \cite{Bellomo1994}, i.e., they resembled the modes localized on POs of the undeformed triangles. But even though their modes retained some of the features of the unperturbed triangles, their properties like the major emission directions could only be explained qualitatively, if at all. This means that, contrary to expectations, the lasing modes of triangular microlasers are not based on a PO if no simple PO exists in the corresponding billiard. It also demonstrates that there seems to be a significant, qualitative difference between microlasers with the shape of rational and irrational triangles which is surprising since any irrational triangle can be approximated by a rational one.

The examples of irrational triangles demonstrate that the predictive power of ray optics decreases for irrational triangles like the Pythagorean triangle, and quantitative predictions are no longer possible for irrational triangles without symmetries and simple POs. There is apparently a transition from symmetric triangles with simple classical dynamics for which the lasing properties can be very well explained by ray optics to less symmetric triangles with increasingly complicated classical dynamics for which ray-optical methods no longer yield a good description of the lasing properties. So the complexity of the classical dynamics is directly reflected in the complexity of the corresponding microlasers. It remains an interesting challenge to better understand the latter cases by further experimental and numerical studies.

\begin{acknowledgments}
The authors thank E.~Bogomolny and A.~Nosich for illuminating and fruitful discussions. S.~B.\ gratefully acknowledges funding from the European Union Seventh Framework Programme (FP7/2007-2013) under Grant No.\ 246.556.10. This work was supported by a public grant from the Laboratoire d'Excellence Physics Atom Light Matter (LabEx PALM) overseen by the French National Research Agency (ANR) as part of the Investissements d'Avenir program (Reference No.\ ANR-10-LABX-0039).
\end{acknowledgments}

\end{document}